\documentstyle[aps]{revtex}
\begin{document}
\bibliographystyle{unsrt}
\draft
\title{Sub-femtosecond determination of transmission delay times for a
dielectric mirror
 (photonic bandgap) as a function of angle of incidence}
\author{Aephraim M. Steinberg\footnote{As of 1995, at National
Institute of Standards and Technology, Phys A167,
Gaithersburg, MD 20899} and Raymond Y. Chiao}
\address{Department of Physics, U.C. Berkeley, Berkeley, CA 94720\\
Internet: aephraim@physics.berkeley.edu}
\date{Preprint quant-ph/9501013; Received Phys. Rev. 17 August, 1994}
\maketitle

\begin{abstract}
Using a two-photon interference technique, we measure the delay for
single-photon wavepackets to be transmitted through a multilayer dielectric
mirror, which functions as a ``photonic bandgap'' medium.  By varying the
angle of incidence, we are able to confirm the behavior predicted
by the group delay (stationary phase approximation), including a variation
of the delay time from superluminal to subluminal as the band edge is tuned
towards to the wavelength of our photons.  The agreement with theory
is better than 0.5 femtoseconds (less than one quarter of an optical
period) except at large angles of incidence.
The source of the remaining discrepancy is not yet fully understood.
\end{abstract}
\pacs{PACS numbers: 03.65.Bz, 42.50.Wm, 73.40.Gk}

\newcommand{\um}[1]{\"{#1}}

In recent years, there has been a great deal of interest in two related topics:
tunneling times
\cite{Buttiker=1982,Hauge=1989,Buttiker=1990,Landauer=1994} and
photonic bandgaps
\cite{Yablonovitch=1993,John=1991,de_Sterke=1988}.  A standard
quarter-wave-stack dielectric mirror is in fact the simplest example of a
one-dimensional photonic bandgap, and consequently may be thought of as
a tunnel barrier for photons within its ``stop band.''  The periodic
modulation of the refractive index is analogous to a periodic Kronig-Penney
potential in solid-state physics, and leads to an imaginary value for the
quasimomentum in certain frequency ranges-- that is, to an exponentially
decaying field envelope within the medium, and high reflectivity due to
constructive interference (Bragg reflection).  We have exploited this analogy
to perform the first measurement of the single-photon tunneling delay time
\cite{Steinberg=1993PRL,Chiao=1993SA,Landauer=1993,%
Stovneng=1993}, using as our barrier an 11-layer mirror of alternating high
(n=2.22) and low (n=1.41) index quarter-wave layers, with minimum
transmission of about 1\% at the center of the bandgap.  We confirmed the
striking prediction which drives the tunneling time controversy: in certain
limits, a transmitted wave packet peak may appear on the far side of the
barrier {\it faster} than if the peak had traversed the barrier at the vacuum
speed of light $c$.  Meanwhile, several microwave experiments have
investigated other instances of superluminal propagation including
electromagnetic analogies to tunneling
\cite{Ishii=1991,Ranfagni=1991APL,%
Enders=1993,Ranfagni=1993,Nimtz=1994,Steinberg=1994COM}.

	While in itself, this anomalous peak propagation does not constitute
a violation of Einstein causality
\cite{Brillouin=1960,Garrett=1970,Chu=1982a,Chu=1982b,%
Bolda=1994PROP,Deutch=1993,%
Japha=1994,Kurizki=1994,Hass=1994,Azbel=1994}, it certainly leads one
to ask whether there may exist another, longer timescale in tunneling, with
more physical significance than the group ({\it i.e.}, peak) delay.  After all,
in a certain sense, the bulk of the transmitted wave originates in the leading
edge of the incident wave packet, {\it not} near the incident peak
\cite{Leavens=1993BK,Deutch=1993,Steinberg=1994HI}.  Many theories
have been propounded to describe the duration of the tunneling interaction,
and the leading contenders involve studying oscillating barriers
\cite{Buttiker=1982,Martin=1993} or Larmor precession of a tunneling
electron in a barrier with a confined magnetic field
\cite{Baz=1967,Rybachenko=1967,Buttiker=1983}.  It should be stressed
that these theories are not intended to describe the propagation of wave
packets, but rather the dynamical timescale of the tunneling process; several
experiments have supported their predictions \cite{Landauer=1989}.

	Nevertheless, there is a popular misconception that these times (and
in particular the B\"uttiker-Landauer time in its ``semiclassical'' or WKB
limit-- $md/\hbar\kappa$, where $\kappa$ represents the evanescent decay
constant inside the barrier, {\it i.e.}, the magnitude of the imaginary
wavevector) predict the arrival time of wave packets.  In
\cite{Steinberg=1993PRL}, we were able to exclude the semiclassical time,
but not B\"uttiker's version of the Larmor time, as describing peak
propagation.  Furthermore, some workers have expressed concern about the
paucity of data supporting the superluminality of the group delay, in spite of
our finding of a seven-standard-deviation effect.  Microwave experiments
have also traditionally been met with skepticism (see, for example,
\cite{Stephan=1993}.)  In light of these objections, we have extended the
earlier experiment to study the delay time as a function of angle-of-%
incidence.  As the angle is changed, the frequency and the width of the
bandgap change as well, so this is essentially a way to study the energy-%
dependence of the tunneling time.

	Our apparatus is shown in Fig. 1.  As the technique and the sample
have both been described at length elsewhere
\cite{Hong=1987,Steinberg=1992PRA,Steinberg=1993PRL,%
Jeffers=1993,Shapiro=1994}, we will content ourselves with an
abbreviated sketch of the method.  A crystal with an optical $\chi^{(2)}$
nonlinearity is pumped by a cw ultraviolet laser, and in the process of
spontaneous parametric down-conversion emits simultaneous pairs of
horizontally-polarized infrared photons.  The two photons in each pair leave
the crystal on opposite sides of the ultraviolet pump, conserving
momentum.  They are correlated in time to within their reciprocal bandwidth
of about 15 fs.  They are also correlated in energy, their frequencies
summing to that of the (narrow-band) 351 nm pump.  When the two
photons arrive simultaneously at a beam splitter, there is no way to
distinguish the two Feynman paths leading to coincidences between
detectors placed at the beam splitter's two output ports: both photons being
transmitted, and both photons being reflected.  This leads to an interference
effect in which the coincidence rate is suppressed (the two photons tending to
head off to the same detector).  By contrast, if the photons arrive at the beam
splitter at different times (on the scale of their 15 fs correlation time),
coincidence counts occur half the time.  Thus by placing a dielectric mirror
in one arm of the interferometer and adjusting the external path length to
minimize the coincidence rate, we can measure the delay experienced by the
photon wavepackets which are transmitted through this barrier.  We find
that near the transmission minimum, the photons travel through the mirror
faster than they travel through an equivalent length of air, whereas when the
mirror is angled to bring the band-edge closer to the photons' wavelength,
they travel slower than through air, as one would expect.  Figure 2 shows
sample data for these two situations, where the sign change can be clearly
seen.

	In \cite{Steinberg=1993PRL}, our results were consistent with the
group delay predictions, and also with B\"uttiker's proposed Larmor time
\cite{Buttiker=1983}, but not with the ``semiclassical'' time.  The measured
times exceeded the predictions by approximately 0.5 fs, but this result was
at the borderline of statistical significance, and not discussed.  Since then,
further data taken at various angles of incidence have continued to show
a discrepancy, ranging from an excess of 0.5 fs near normal incidence to
a deficit of over 1 fs at large angles of incidence.  At the same time,
the data offer close agreement
with the group delay, and appear to rule out identification of the Larmor
theory with a peak propagation time.
Our attempts to eliminate
systematic effects and characterize those which remain were described in
\cite{Steinberg=1993PRL}.  Since then, unable to find any other sources of
error to explain the discrepancy, we are convinced that it is a property of the
sample under study, and not of the interferometer used for the measurements.
We therefore obtained a second dielectric mirror of design parameters
identical to the first, to see whether the errors could be attributed
to deviations from the ideal quarter-wave-stack structure.
As can be seen from Figure
3, both mirrors show quite similar behavior.
Both are 11-layer
quarter-wave stacks as described above.  Mirror 1 shows a minimum
transmission at 692 nm, while mirror 2's minimum is at 688 nm; this
difference is insignificant on the scale of the bandgap, which extends from
600 nm to 800 nm.  We conclude that some
real effect is at work, modifying the stationary-phase
prediction.  In principle, frequency-dependent transmission could lead to
such an effect, as does second-order group-velocity dispersion; both effects
are much too small to explain the present discrepancy.  As discussed in
\cite{Steinberg=1994TH}, attempts to numerically model dielectric mirrors
with small, random fluctuations in layer thicknesses were able to produce
deviations on the right order, but in general they did not lead to
deviations of the form we observed experimentally.
It is conceivable that loss or scattering in the
dielectrics could also help
explain the effect, and we are beginning to investigate this
possibility \cite{Steinberg=1994PRA}; see also
\cite{Ranfagni=1990,Mugnai=1992}.

	Theoretical curves are plotted along with the data in figures
3 and 4.  The group delay is calculated by the method of stationary
phase.  The transmission phase of the 11-layer structure is calculated
numerically, and differentiated first with respect to angle of incidence
to give the transverse shift
and then with respect to incident frequency
to give the time delay, according to the formulas $\Delta y =
-\partial\phi_T/\partial k_y =
-\partial\phi_T/\partial(k\sin\theta) = -(k\cos\theta)
\partial\phi_T/\partial\theta$
and $\tau_g = \partial\phi_T/\partial\omega + (\Delta y/c)\sin\theta$,
where $\phi_T$ is the transmission phase \cite{Steinberg=19941D2D}.
B\"uttiker's Larmor time \cite{Buttiker=1983}
is equal to the magnitude of the complex time \cite{Sokolovski=1993,%
Steinberg=1994WK}
$\tau_c = i \partial ({\rm ln}\;t) / \partial \Omega_L$, where $t$ is
the complex transmission amplitude, and $\Omega_L$ the Larmor frequency.
For our optical structure, an effective Larmor frequency $\Omega_L$
corresponds to a uniform (over the barrier region) scaling of the local
index of refraction by a factor of $1+\Omega_L/\omega$.
Since in the limit of interest, the ``in-plane portion'' of the Larmor
time ({\it i.e.}, the real part of the complex time) differs little from the
group delay, we take them to be equal in order to include the effects of
the transverse shift in the Larmor theory.  The ``out-of-plane portion''
(or imaginary part) is calculated numerically, and added in quadrature
to the group delay in order to generate the Larmor time.
Since our measurements compare the transit time through the barrier
with that through air, we subtract the time parallel wavefronts propagating
at $c$ would take to reach a point on the far side of the barrier (with a
transverse shift of $\Delta y$) from both the group delay and the
Larmor time, so as to facilitate comparison with the experimental data.

	At the moment, more work (both experimental and theoretical) is
needed to understand the discrepancy.  We are therefore planning to repeat
this experiment with s-polarized light (by introducing half-wave plates
before and after the sample being studied), which has very different
transmission characteristics as a function of angle (also leading to a larger
difference between the group delay and the Larmor theories).  As shown in
Figure 4, our preliminary data are again consistent with the group delay and
not with the Larmor time, but due to the lower transmission for this
polarization, we need to improve our signal-to-noise ratio before reaching any
definitive conclusions.

The superluminality of the barrier traversal near mid-gap is now well
supported by the data, and the group delay (stationary-phase) theory can be
seen to be relatively accurate for a variety of angles of incidence, but there
is a residual discrepancy on the order of 0.5 femtoseconds, which is not
yet fully understood.

This work was supported by the U.S. Office of Naval Research
under grant N00014-90-J-1259.
We would like to thank P. G. Kwiat, M. Mitchell, B. Johnson, G.
McKinney, and J. Holden for their assistance; W. Heitmann for
discussions about data analysis; and R. Landauer, J. D.
Jackson, and E. Yablonovitch for useful disagreements.

Notes added during revision: Since the submission of this manuscript,
a paper has appeared \cite{Spielmann=1994}
extending our previous experimental results to barriers
of varying thicknesses (and transmission as low as $0.01 \%$) near
normal incidence, using classical femtosecond pulses.  It reports
general agreement with the group delay theory, aside from a discrepancy
on the order of 1.5 fs.  Two papers have also appeared discussing the
effects of dissipation on tunneling times \cite{Nimtz=1994DIS,%
Raciti=1994}.

% \references

\newpage
\renewcommand{\theenumii}{\alph{enumii}}

\begin{center}
{\bf Figure Captions}
\end{center}
\begin{enumerate}
\item Experimental setup for determining single-photon propagation times
through a multilayer dielectric mirror.  By translating the sample, we
can observe the intereference dip for photons tunneling through the
$1.1 \mu m$ barrier or for photons traversing an equal thickness of air.
We can thus compare arrival times for tunneling and freely propagating
wave packets.

\item Coincidence rate versus trombone prism position (see Fig. 1) for
p-polarized photons travelling through the reflective coating as well as for
those travelling through an equal thickness of air, for (a) normal incidence
and (b) 55-degree incidence.

\item Left axis: measured delay for mirror one (squares) and mirror two
(circles) as a function of angle of incidence, to be compared with the
theoretical group delay and the Larmor interaction time proposed by
B\"uttiker.  Right axis: transmission versus angle of incidence.  All curves
are for p-polarization.

\item Same as Fig. 3, but for s-polarization.  Due to the much lower
transmission, only one preliminary data point is shown, but the different
characteristics of the theoretical curves for both transmission and delay
suggest that upon improvement of our signal-to-noise ratio, further work in
this direction may help elucidate the discrepancy between experiment and
theory.

\end{enumerate}


\begin{thebibliography}{10}

\bibitem{Buttiker=1982}
M. B\um{u}ttiker and R. Landauer, Phys. Rev. Lett. {\bf 49}, 1739 (1982).

\bibitem{Hauge=1989}
E. H. Hauge and J. A. St{\o}vneng, Rev. Mod. Phys. {\bf 61}, 917 (1989).

\bibitem{Buttiker=1990}
M. B\um{u}ttiker, {\it In} ``Electronic Properties of Multilayers and Low
  Dimensional Semiconductors,'' J. M. Chamberlain, L. Eaves, and J. C. Portal,
  eds., Plenum, New York (1990).

\bibitem{Landauer=1994}
R. Landauer and Th. Martin, Rev. Mod. Phys. {\bf 66}, 217 (1994).

\bibitem{Yablonovitch=1993}
E. Yablonovitch, J. Opt. Soc. Am. B {\bf 10} (2), 283 (1993), and references
  therein, and other articles in this special issue on photonic bandgaps.

\bibitem{John=1991}
S. John, Physics Today {\bf 44}, 32 (1991).

\bibitem{de_Sterke=1988}
C. M. de Sterke and J. E. Sipe, Phys. Rev. A {\bf 38}, 5149 (1988).

\bibitem{Steinberg=1993PRL}
A. M. Steinberg, P. G. Kwiat, and R. Y. Chiao, Phys. Rev. Lett. {\bf 71}, 708
  (1993).

\bibitem{Chiao=1993SA}
R.Y. Chiao, P.G. Kwiat, and A.M. Steinberg, ``Faster Than Light?'' {\it
  Scientific American} {\bf 269}, 52 (1993).

\bibitem{Landauer=1993}
R. Landauer, Nature {\bf 365}, 692 (1993).

\bibitem{Stovneng=1993}
J. A. St{\o}vneng and E. H. Hauge, Phys. World {\bf 6}, 23 (1993).

\bibitem{Ishii=1991}
T.K. Ishii and G.C. Giakos, Microwaves and RF {\bf 30}, 114 (1991) and
  references therein.

\bibitem{Ranfagni=1991APL}
A. Ranfagni, D. Mugnai, P. Fabeni, and G. P. Pazzi, Appl.Phys.Lett. {\bf 58},
  774 (1991).

\bibitem{Enders=1993}
A. Enders and G. Nimtz, J. Phys. I France {\bf 3}, 1089 (1993).

\bibitem{Ranfagni=1993}
A. Ranfagni, P. Fabeni, G.P. Pazzi, and D. Mugnai, Phys. Rev. E {\bf 48}, 1453
  (1993).

\bibitem{Nimtz=1994}
G. Nimtz, A. Enders, and H. Spieker, J. Phys. I France {\bf 4}, 565 (1994).

\bibitem{Steinberg=1994COM}
A. M. Steinberg, J. Phys I France {\bf 4}, 1813 (1994).

\bibitem{Brillouin=1960}
L. Brillouin. {\it Wave Propagation and Group Velocity}. Academic, New York
  (1960).

\bibitem{Garrett=1970}
C. G. B. Garrett and D. E. McCumber, Phys. Rev. A {\bf 1}, 305 (1970).

\bibitem{Chu=1982a}
S. Chu and S. Wong, Phys. Rev. Lett. {\bf 48}, 738 (1982).

\bibitem{Chu=1982b}
S. Chu and S. Wong, Phys. Rev. Lett. {\bf 49},1292 (1982).

\bibitem{Bolda=1994PROP}
E. L. Bolda, J. C. Garrison, and R. Y. Chiao, Phys. Rev. A {\bf 49}, 2938
  (1994).

\bibitem{Deutch=1993}
J. M. Deutch and F. E. Low, Ann. Phys. {\bf 228}, 184 (1993).

\bibitem{Japha=1994}
Y. Japha and G. Kurizki, ``Unified Causal Theory of Photonic Superluminal
  Delays in Dielectric Structures,'' submitted to Phys. Rev. Lett., 1994.

\bibitem{Kurizki=1994}
G. Kurizki and Y. Japha, ``Causality and Interference in Electron and Photon
  Tunneling,'' to appear in {\it Quantum Interferometry}, proceedings of the
  Adriatico Research Workshop on Quantum Interferometry (Trieste, March 1993),
  organized by F. de Martini and A. Zeilinger.

\bibitem{Hass=1994}
K. Hass and P. Busch, Phys. Lett. A {\bf 185}, 9 (1994).

\bibitem{Azbel=1994}
Mark Ya. Azbel, Sol. St. Comm. {\bf 91}, 439 (1994).

\bibitem{Leavens=1993BK}
C. R. Leavens and G. C. Aers. {\it In} ``Scanning Tunneling Microscopy III,''
  R. Wiesendanger and H.-J. G\um{u}ntherodt, eds., Springer-Verlag, Berlin
  (1993).

\bibitem{Steinberg=1994HI}
A. M. Steinberg, P. G. Kwiat, and R. Y. Chiao, Found. Phys. Lett. {\bf 7}, 223
  (1994).

\bibitem{Martin=1993}
T. Martin and R. Landauer, Phys. Rev. A {\bf 47}, 2023 (1993).

\bibitem{Baz=1967}
A. I. Baz', Sov. J. Nucl. Phys. {\bf 5}, 161 (1967).

\bibitem{Rybachenko=1967}
V. F. Rybachenko, Sov. J. Nucl. Phys. {\bf 5}, 635 (1967).

\bibitem{Buttiker=1983}
M. B\um{u}ttiker, Phys. Rev. {\bf B27}, 6178 (1983).

\bibitem{Landauer=1989}
R. Landauer, Nature {\bf 341}, 567 (1989).

\bibitem{Stephan=1993}
K.D. Stephan, IEEE Antennas and Propagation Magazine {\bf 35}, 13 (1993).

\bibitem{Hong=1987}
C. K. Hong, Z. Y. Ou, and L. Mandel, Phys. Rev. Lett. {\bf 59}, 2044 (1987).

\bibitem{Steinberg=1992PRA}
A. M. Steinberg, P. G. Kwiat, and R. Y. Chiao, Phys. Rev. {\bf A45}, 6659
  (1992).

\bibitem{Jeffers=1993}
J. Jeffers and S. M. Barnett, Phys. Rev. A, {\bf 47}, 3291 (1993).

\bibitem{Shapiro=1994}
J. H. Shapiro and K.-X. Sun, J. Opt. Soc. Am. B {\bf 11}, 1130 (1994).

\bibitem{Steinberg=1994TH}
A. M. Steinberg, {\it When Can Light Go Faster Than Light? The tunneling time
  and its sub-femtosecond measurement via quantum interference}, {\it Ph. D.}
  thesis, University of California at Berkeley, 1994.

\bibitem{Steinberg=1994PRA}
A. M. Steinberg, ``Conditional probabilities in quantum theory, and the
  tunneling time controversy,'' submitted to Phys. Rev. A (1994).

\bibitem{Ranfagni=1990}
A. Ranfagni, D. Mugnai, P. Fabeni, and G. P. Pazzi, Physica Scripta {\bf 42},
  508 (1990).

\bibitem{Mugnai=1992}
D. Mugnai, A. Ranfagni, R. Ruggeri, and A. Agresti, Phys. Rev. Lett. {\bf 68},
  259 (1992).

\bibitem{Steinberg=19941D2D}
A. M. Steinberg and R. Y. Chiao, Phys. Rev. A {\bf 49}, 3283 (1994).

\bibitem{Sokolovski=1993}
D. Sokolovski and J. N. L. Connor, Phys. Rev. A {\bf 47}, 4677 (1993).

\bibitem{Steinberg=1994WK}
A. M. Steinberg, ``How much time does a tunneling particle spend in the barrier
  region?'', to appear in Phys. Rev. Lett. (1995).

\bibitem{Spielmann=1994}
Ch. Spielmann, R. Szip\um{o}cs, A. Stingl, and F. Krausz, Phys. Rev. Lett. {\bf
  73}, 2308 (1994).

\bibitem{Nimtz=1994DIS}
G. Nimtz, H. Spieker, and H.M. Brodowsky, J. Phys. I France {\bf 4}, 1379
  (1994).

\bibitem{Raciti=1994}
F. Raciti and G. Salesi, J. Phys. I France {\bf 4}, 1783 (1994).

\end{thebibliography}
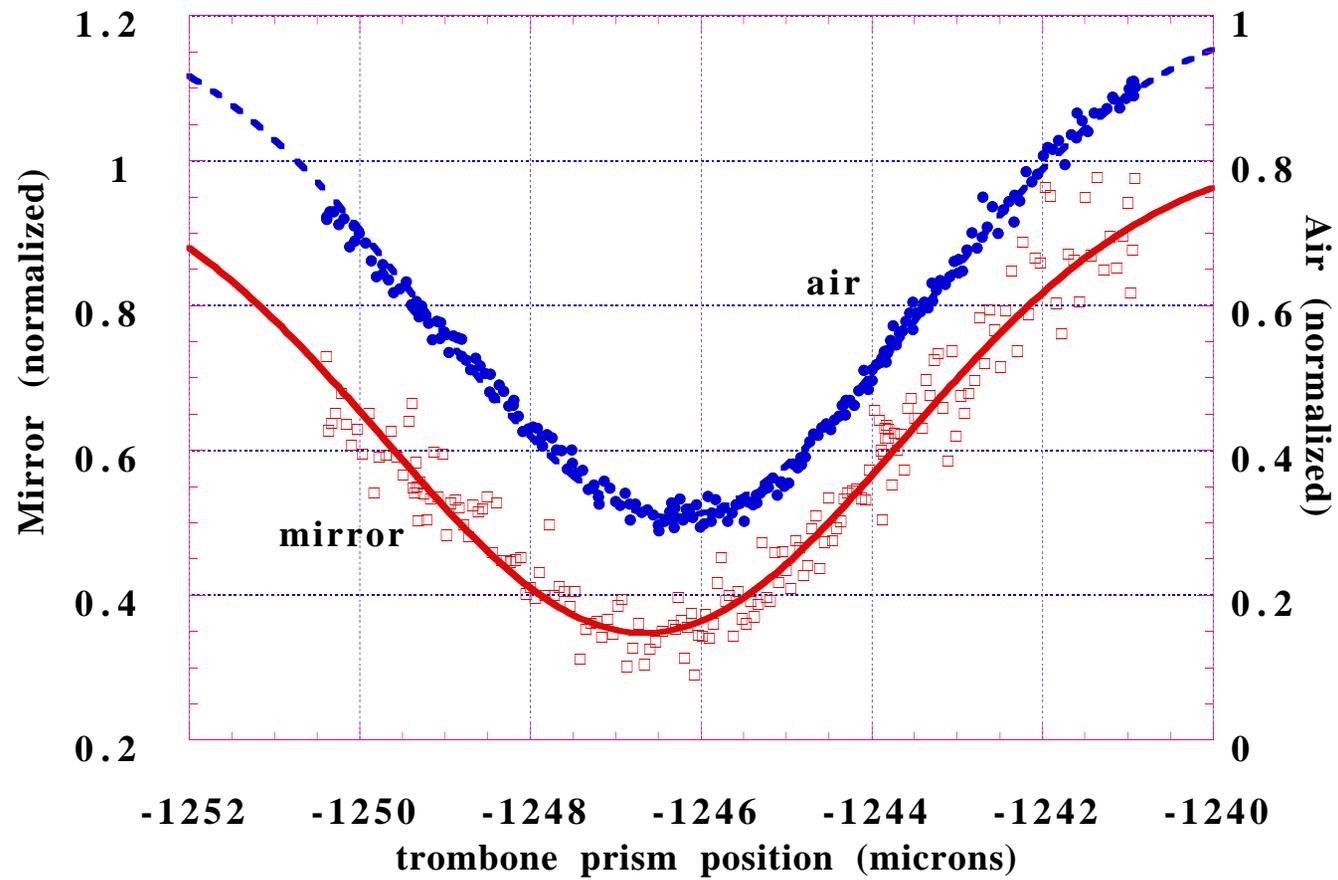
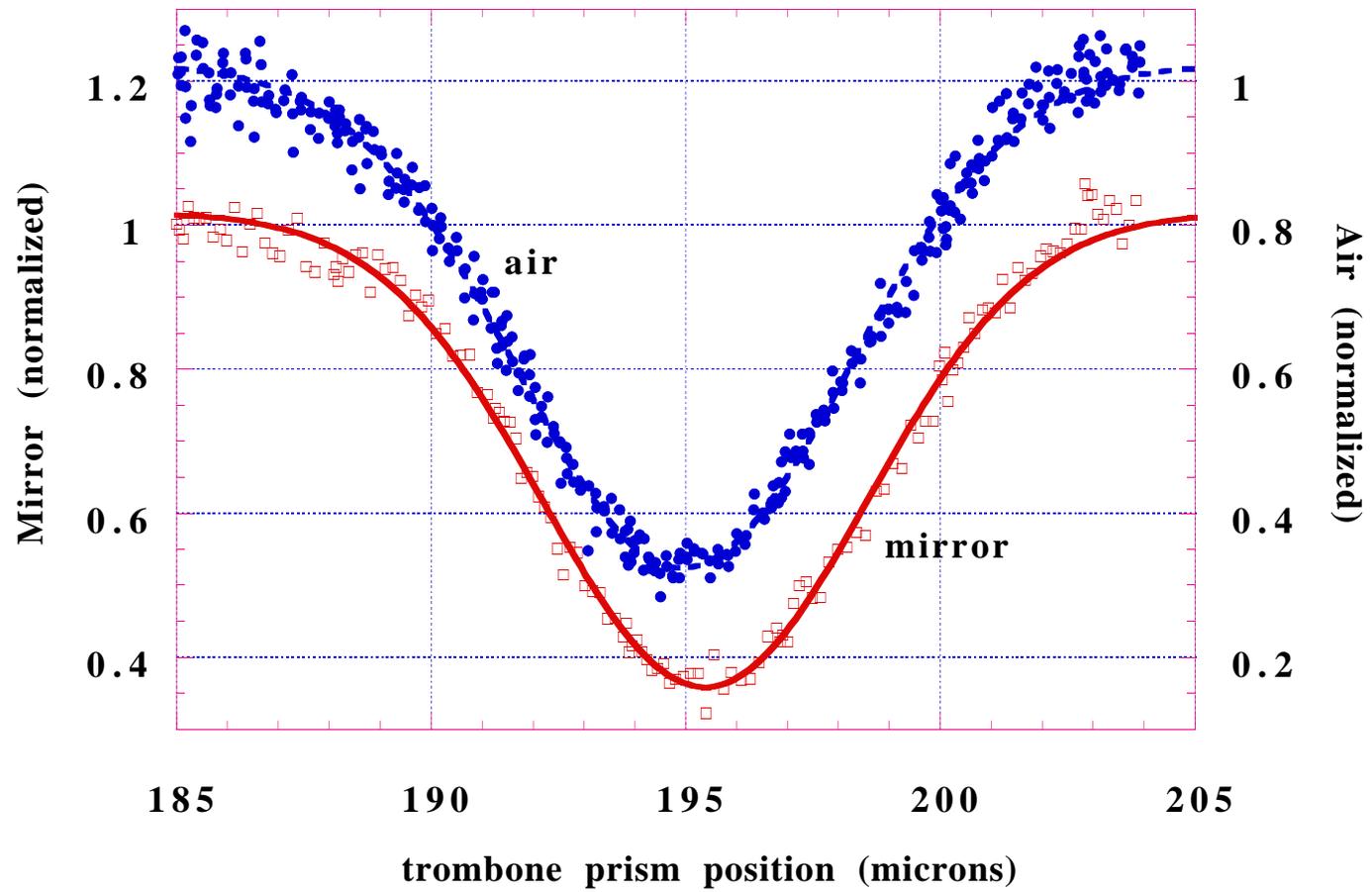
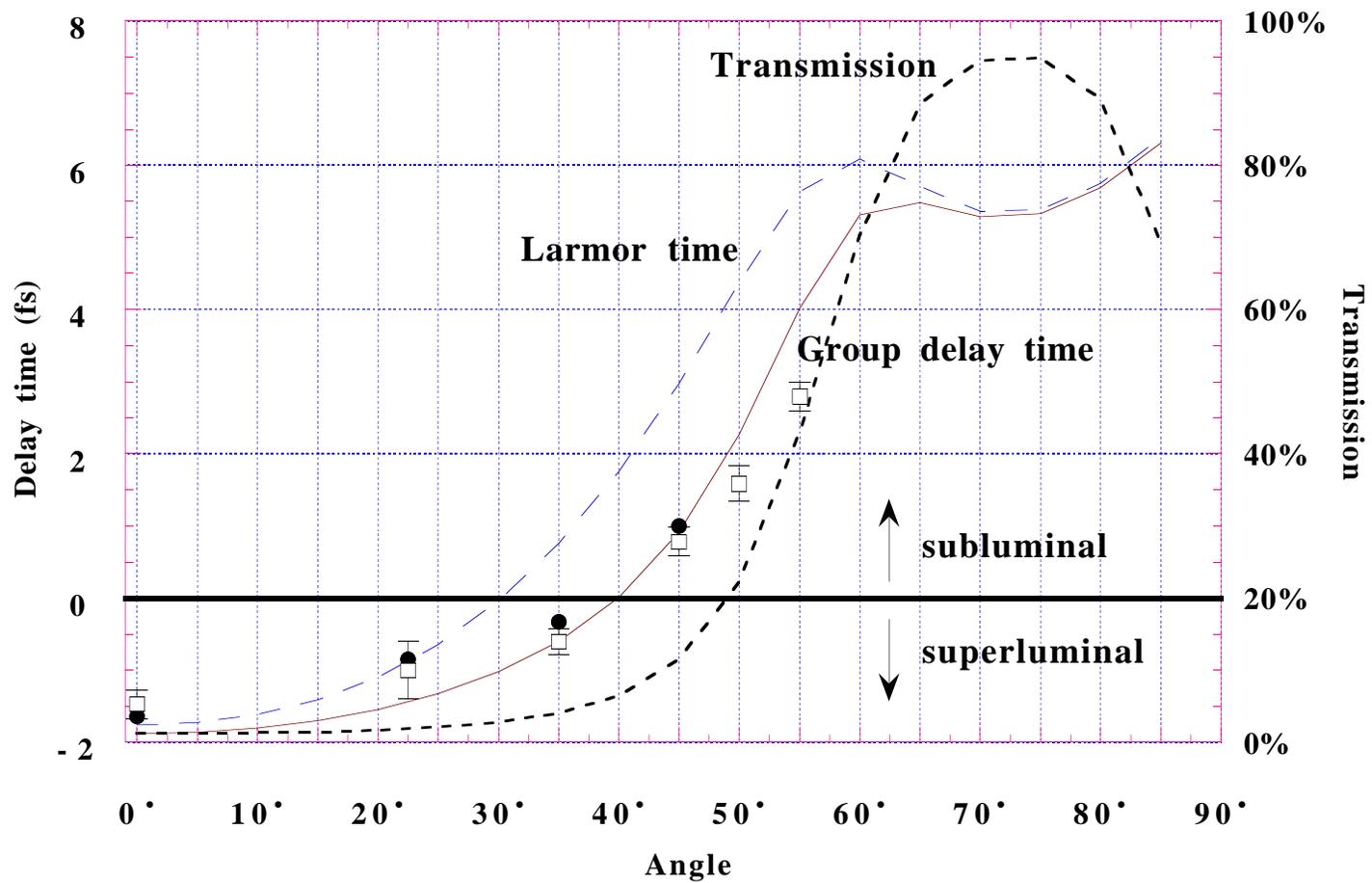
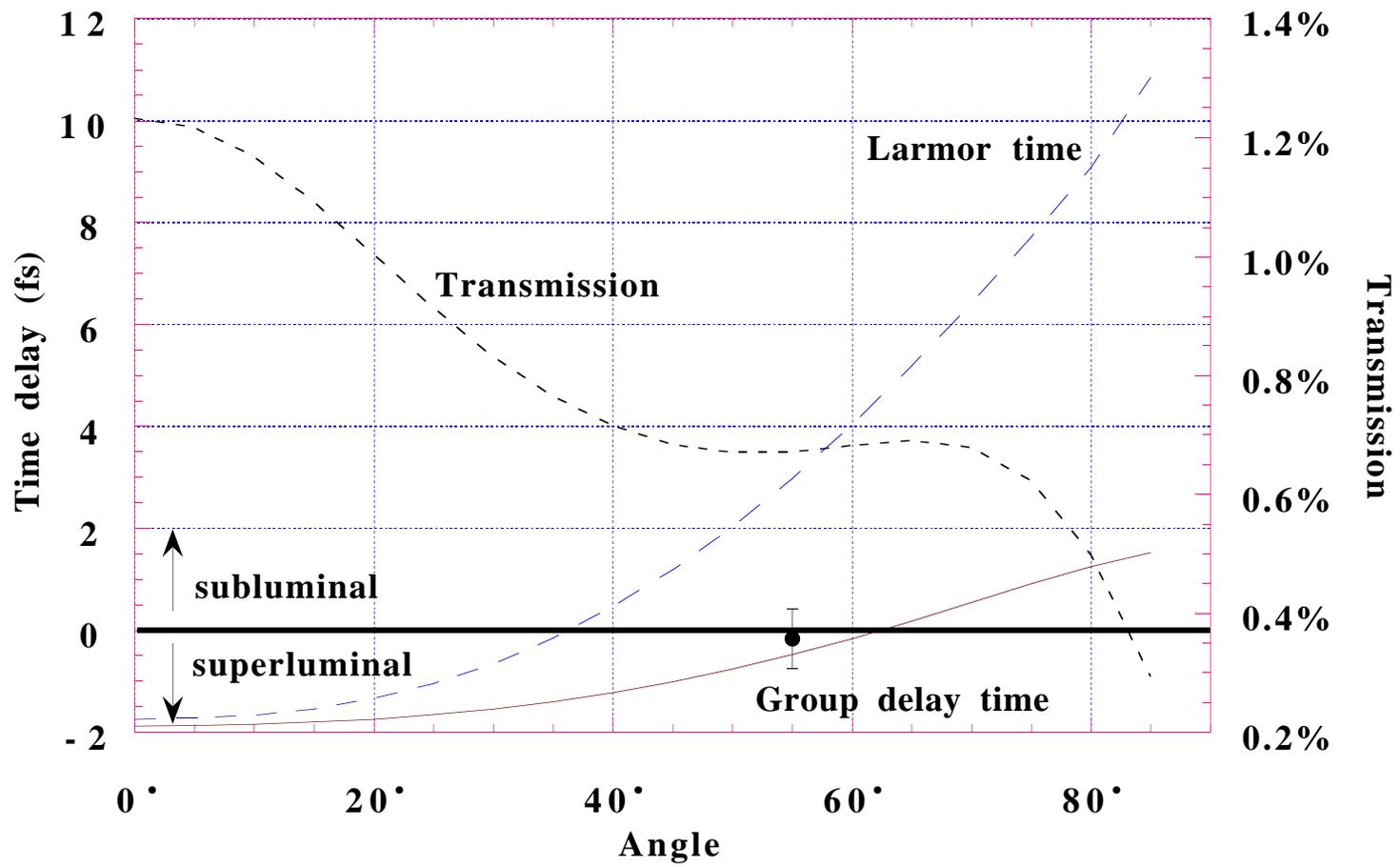
\end{document}